# Intricate Knots in Proteins: Function and Evolution

Peter Virnau[1]*, Leonid A. Mirny[1,2], Mehran Kardar[1]

1 Department of Physics, Massachusetts Institute of Technology, Cambridge, Massachusetts, United States of America, 2 Harvard–MIT Division of Health Sciences and Technology, Cambridge, Massachusetts, United States of America

Our investigation of knotted structures in the Protein Data Bank reveals the most complicated knot discovered to date. We suggest that the occurrence of this knot in a human ubiquitin hydrolase might be related to the role of the enzyme in protein degradation. While knots are usually preserved among homologues, we also identify an exception in a transcarbamylase. This allows us to exemplify the function of knots in proteins and to suggest how they may have been created.

## Introduction

Although knots are abundant and complex in globular homopolymers [1–3], they are rare and simple in proteins [4–8]. Sixteen methyltransferases in bacteria and viruses can be combined into the α/β knot superfamily [9], and several isozymes of carbonic anhydrase (I, II, IV, V) are known to be knotted. Apart from these two folds, only a few insular knots have been reported [5,6,10,11], some of which were derived from incomplete structures [6,11]. For the most part, knotted proteins contain simple trefoil knots ($3_1$) that can be represented by three essential crossings in a projection onto a plane (see Figure 1, left). Only three proteins were identified with four projected crossings ($4_1$, Figure 1, middle).

In this report we provide the first comprehensive review of knots in proteins, which considers all entries in the Protein Data Bank (http://www.pdb.org) [12], and not just a subset. This allows us to examine knots in homologous proteins. Our analysis reveals several new knots, all in enzymes. In particular, we discovered the most complicated knot found to date ($5_2$) in human ubiquitin hydrolase (Figure 1, right), and suggest that its entangled topology protects it against being pulled into the proteasome. We also noticed that knots are usually preserved among structural homologues. Sequence similarity appears to be a strong indicator for the preservation of topology, although differences between knotted and unknotted structures are sometimes subtle. Interestingly, we have also identified a novel knot in a transcarbamylase that is not present in homologues of known structure. We show that the presence of this knot alters the functionality of the protein, and suggest how the knot may have been created in the first place.

Mathematically, knots are rigorously defined in closed loops [13]. Fortunately, both the N- and C-termini of open proteins are typically accessible from the surface and can be connected unambiguously: we reduce the protein to its $C_\alpha$-backbone, and draw two lines outward starting at the termini in the direction of the connection line between the center of mass of the backbone and the respective ends [5]. The lines are joined by a big loop, and the structure is topologically classified by the determination of its Alexander polynomial [1,13]. Applying this method to the Protein Data Bank in the version of January 3, 2006, we found 273 knotted structures in the 32,853 entries that contain proteins (Table S1). Knots formed by disulfide [14,15] or hydrogen bonds [7] were not included in the study.

## Results

For further analysis, we considered 36 proteins that contain knots as defined by rather stringent criteria discussed in the Materials and Methods section. These proteins can be classified into six distinct families (Table 1). Four of these families incorporate a deeply knotted section, which persists when 25 amino acids are cut off from either terminus. Interestingly, all knotted proteins thus identified are enzymes. Our investigation affirms that all members of the carbonic anhydrase fold (including the previously undetermined isozymes III, VII, and XIV) are knotted. In addition, we identify a novel trefoil in two bacterial transcarbamylase-like proteins (AOTCase in *Xanthomonas campestris* and SOTCase in *Bacteroides fragilis*) [16,17].

**UCH-L3—The most complex protein knot.** One of our most intriguing discoveries is a fairly intricate knot with five projected crossings ($5_2$) in ubiquitin hydrolase (*UCH-L3* [18]; see Figure 1, right). This knot is the first of its kind and, apart from carbonic anhydrases, the only identified in a human protein. Human *UCH-L3* also has a yeast homologue [6,19] with a sequence identity of 32% [20]. Amino acids 63 to 77 are unstructured, and if we connect the unstructured region by an arc that is present in the human structure, we obtain the same knot with five crossings. What may be the function of this knot? In eukaryotes, proteins get labeled for Abbreviations: AOTCase, N-acetylornithine transcarbamylase; SOTCase, N-succinylornithine transcarbamylase; UCH-L3, ubiquitin hydrolase

* To whom correspondence should be addressed. E-mail: virnau@mit.edu

## Synopsis

Several protein structures incorporate a rather unusual structural feature: a knot in the polypeptide backbone. These knots are extremely rare, but their occurrence is likely connected to protein function in as yet unexplored fashion. The authors' analysis of the complete Protein Data Bank reveals several new knots that, along with previously discovered ones, may shed light on such connections. In particular, they identify the most complex knot discovered to date in a human protein, and suggest that its entangled topology protects it against unfolding and degradation. Knots in proteins are typically preserved across species and sometimes even across kingdoms. However, there is also one example of a knot in a protein that is not present in a closely related structure. The emergence of this particular knot is accompanied by a shift in the enzymatic function of the protein. It is suggested that the simple insertion of a short DNA fragment into the gene may suffice to cause this alteration of structure and function.

degradation by ubiquitin conjugation. *UCH-L3* performs deconjugation of ubiquitin, thus rescuing proteins from degradation. The close association of the enzyme with ubiquitin should make it a prime target for degradation at the proteasome. We suggest that the knotted structure of *UCH-L3* makes it resistant to degradation. In fact, the first step of protein degradation was shown to be ATP-dependent protein unfolding by threading through a narrow pore (~13 Å in diameter) of a proteasome [21,22]. Such threading into the degradation chamber depends on how easily a protein unfolds, with more stable proteins being released back into solution [23] and unstable ones being degraded. If ATP-dependent unfolding proceeds by pulling the C-terminus into a narrow pore [21], then a knot can sterically preclude such translocation, hence preventing protein unfolding and degradation. While arceabacterial proteasome PAN was shown to process proteins from its C- to N-terminus [21], it cannot be ruled out that some eukaryotic proteasomes process proteins in the N- to C-direction, thus requiring protection of both termini. Unfolding of a knotted protein by pulling may require a long time for global unfolding and untangling of the knot. Unknotted proteins, in contrast, have been shown to become unstable if a few residues are removed from their termini [24], suggesting that threading a few (5–10) residues into a proteasomal pore would be sufficient to unravel an unknotted structure. At both termini, *UCH-L3* contains loops entangled into the knot protecting both ends against unfolding if pulled. It should also be noted that both N- and C-termini are stabilized by a number of hydrophobic interactions with the rest of the protein. The C-terminus is

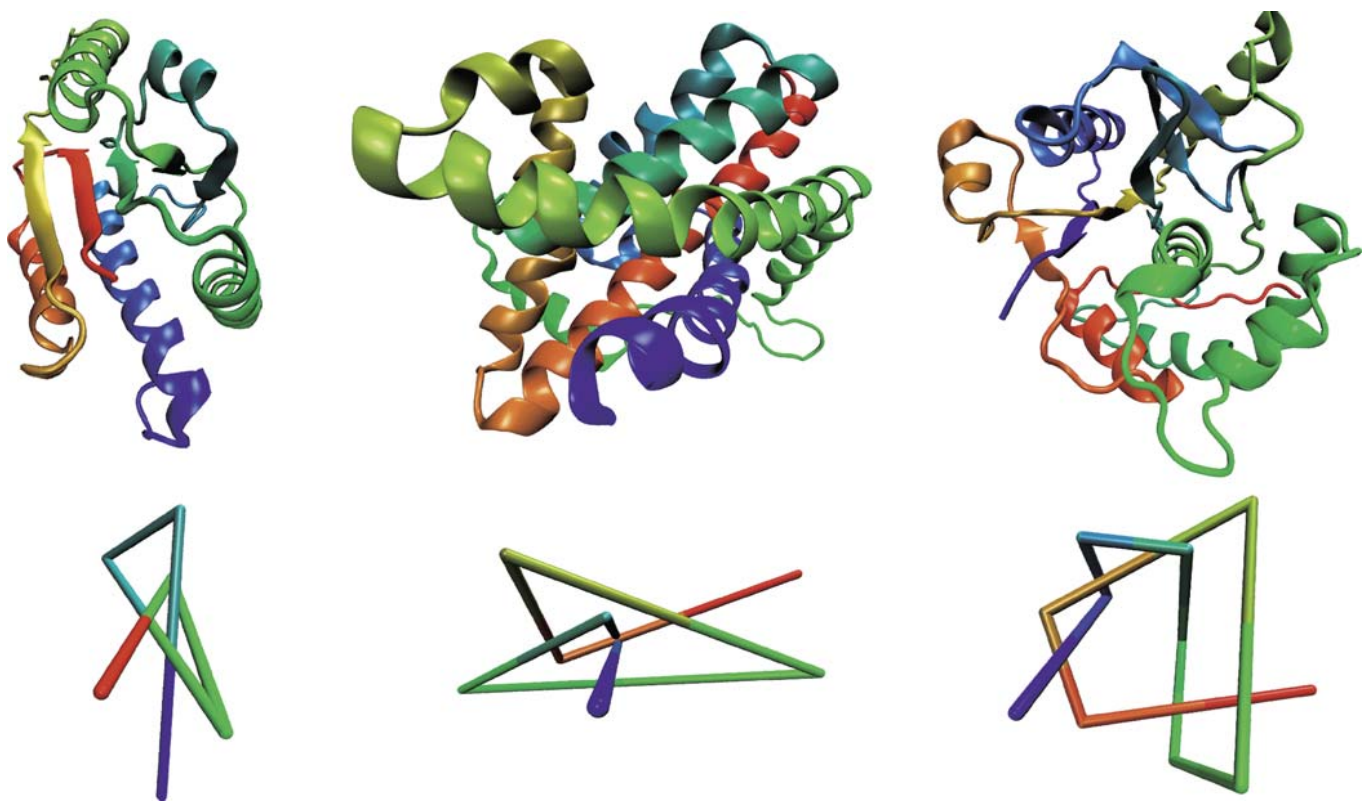

**Figure 1.** Examples of the Three Different Types of Knots Found in Proteins

Colors change continuously from red (first residue) to blue (last residue). A reduced representation of the structure, based on the algorithm described in [1,6,36], is shown in the lower row.
(Left) The trefoil knot ($3_1$) in the YBEA methyltransferase from *E. coli* (pdb code 1ns5; unpublished data) reveals three essential crossings in a projection onto a plane.
(Middle) The figure-eight knot ($4_1$) in the Class II ketol-acid reductoisomerase from *Spinacia oleracea* (pdb code 1yve [26]) features four crossings. (Only the knotted section of the protein is shown.)
(Right) The knot $5_2$ in ubiquitin hydrolase UCH-L3 (pdb code 1xd3 [18]) reveals five crossings. Pictures were generated with Visual Molecular Dynamics (http://www.ks.uiuc.edu/Research/vmd) [43].
DOI: 10.1371/journal.pcbi.0020122.g001

**Table 1.** List of Knotted PDB Entries (January 2006)

| Protein Knot Family | Protein | Species | PDB Code | Length | Knot | Knotted Core |
|---|---|---|---|---|---|---|
| α/β knot | YbeA-like | E. coli | 1ns5 | 153 | $3_1$ | 69–121 (32) |
| | | T. maritime | 1o6d | 147 | $3_1$ | 68–117 (30) |
| | | S. aureus | 1vh0 | 157 | $3_1$ | 73–126 (31) |
| | | B. subtilis | 1to0 | 148 | $3_1$ | 64–116 (32)[a] |
| | tRNA(m1G37)-methyltransferase TrmD | H. influenza | 1uaj | 241 | $3_1$ | 93–138 (92)[a] |
| | | E. coli | 1p9p | 235 | $3_1$ | 90–130 (89)[a] |
| | SpoU-like RNA 2′-O ribose mtf. | T. thermophilus | 1v2x | 191 | $3_1$ | 96–140 (51) |
| | | H. influenza | 1j85 | 156 | $3_1$ | 77–114 (42) |
| | | T. thermophilus | 1ipa | 258 | $3_1$ | 185–229 (29)[a] |
| | | E. coli | 1gz0 | 242 | $3_1$ | 172–214 (28) |
| | | A. aeolicus | 1zjr | 197 | $3_1$ | 95–139 (58) |
| | | S. viridochromog. | 1x7p | 265 | $3_1$ | 192–234 (31) |
| | YggJ C-terminal domain-like | H. influenza | 1nxz | 246 | $3_1$ | 165–216 (30) |
| | | B. subtilis | 1vhk | 235 | $3_1$ | 158–208 (27)[b] |
| | | T. thermophilus | 1v6z | 227 | $3_1$ | 103–202 (25)[c] |
| | Hypothetical protein MTH1 (MT0001) | A. M. Thermoautotr. | 1k3r | 262 | $3_1$ | 48–234 (28) |
| Carbonic anhydrases | Carbonic anhydrase | N. gonorrhoeae | 1kop | 223 | $3_1$ | 36–223 (0) |
| | Carbonic anhydrase I | H. sapiens | 1hcb | 258 | $3_1$ | 29–256 (2) |
| | Carbonic anhydrase II | H. sapiens | 1lug | 259 | $3_1$ | 30–256 (3) |
| | | Bos Taurus | 1v9e | 259 | $3_1$ | 32–256 (3) |
| | | Dunaliella salina | 1y7w | 274 | $3_1$ | 37–270 (4) |
| | Carbonic anhydrase III | Rattus norv. | 1flj | 259 | $3_1$ | 30–256 (3) |
| | | H. sapiens | 1z93 | 263 | $3_1$ | 28–254 (9) |
| | Carbonic anhydrase IV | H.sapiens | 1znc | 262 | $3_1$ | 32–261 (1) |
| | | Mus musculus | 2znc | 249 | $3_1$ | 32–246 (3)[b] |
| | Carbonic anhydrase V | Mus musculus | 1keq | 238 | $3_1$ | 7–234 (4) |
| | Carbonic anhydrase VII | H. sapiens | 1jd0 | 260 | $3_1$ | 28–257 (3) |
| | Carbonic anhydrase XIV | Mus Musculus | 1rj6 | 259 | $3_1$ | 29–257 (2) |
| Miscellaneous | Ubiquitin hydrolase UCH-L3 | H. sapiens | 1xd3A | 229 | $5_2$ | 12–172 (11)[d,e] |
| | | S. cerevisiae (synth.) | 1cmxA | 214 | $3_1$ | 9–208 (6)[b,d] |
| | S-adenosylmethionine synthetase | E. coli | 1fug | 383 | $3_1$ | 33–260 (32) |
| | | Rattus norv. | 1qm4 | 368 | $3_1$ | 30–253 (29)[b] |
| | Class II ketol-acid reductoisomerase | Spinacia oleracea | 1yve | 513 | $4_1$ | 239–451 (62) |
| | | E. coli | 1yrl | 487 | $4_1$ | 220–435 (52) |
| | Transcarbamylase-like | B. fragilis | 1js1 | 324 | $3_1$ | 169–267 (57) |
| | | X. campestris | 1yh1 | 334 | $3_1$ | 171–272 (62) |

"Protein" describes the name or the family of the knotted structure. "Species" refers to the scientific name of the organism from which the protein was taken for structure determination. "PDB code" gives one example Protein Data Bank entry for each knotted protein: additional structures of the same protein can be found using the SCOP classification tool [9]. "Length" describes the number of $C_\alpha$-backbone atoms in the structure. "Knot" refers to the knot type which was discovered in the protein: $3_1$, trefoil; $4_1$, figure-eight knot; $5_2$, 2nd knot with five crossings according to standard knot tables [13]. The core of a knot is the minimum configuration which stays knotted after a series of deletions from each terminus; in brackets we indicate how many amino acids can be removed from either side before the structure becomes unknotted (see Materials and Methods).
[a]Structure is fragmented and becomes knotted when missing sections are joined by straight lines. The size of the knotted core refers to the thus-connected structure. The knot is also present in at least one fragment.
[b]Structure is fragmented and only knotted when missing sections are joined by straight lines. Fragments are unknotted.
[c]1v6z is currently not classified according to SCOP (version 1.69). Sequence similarity suggests that it is part of the α/β knot fold. 1v6zB contains a shallow composite knot ($3_1\#4_1$), which turns into a regular trefoil when two amino acids are cut from the N-terminus. (The random closure [see Materials and Methods] determines a trefoil right away.)
[d]1uch contains the same structure as 1xd3. If the missing section in the center of this structure is joined by a straight line, it becomes knotted ($5_2$). In the yeast homologue (1cmx), amino acids 63 to 77 are unstructured, and if we replace the missing parts by a straight line, we obtain a trefoil knot that has been identified before [6]. If we connect the unstructured region by an arc present in the human structure, we obtain the same knot with five crossings.
[e]Visual inspection reveals that the calculated size of the knotted core is too small.
DOI: 10.1371/journal.pcbi.0020122.t001

particularly stable—residues 223 to 229 are hydrophobic and form numerous contacts at 5 Å with the rest of the structure.

We would like to stress that this hypothesis needs to be tested by experiments. Different proteins may also provide different levels of protection against degradation, depending on structural details, the depth of the knot, and its complexity. Recently, a knot in the red/far-red light photoreceptor phytochrome A in *Deinococcus radiodurans* was identified [11] (see Materials and Methods). Although sequence similarity suggests that the knot may also be present in plant homologues, we cannot be certain. In plants, the red-absorbing form is rather stable (half-life of 1 wk), but the far-red–absorbing form is degraded upon photoconversion by the proteasome with a half-life of 1–2 h in seedlings (and somewhat longer in adult plants) [25].

**Evolutionary aspects.** As expected, homologous structures tend to retain topological features. The trefoil knot in carbonic anhydrase can be found in isozymes ranging from bacteria and algae to humans (Table 1). Class II ketol-acid reductoisomerase comprises a figure-eight knot present in *Escherichia coli* [10] and spinach [26] (see Figure 1, middle), and S-adenosylmethione synthetase contains a deep trefoil knot in *E. coli* [5,27] and rat [28]. It appears that particular knots have indeed been preserved throughout evolution, which suggests a crucial role for knots in protein enzymatic activity and binding.

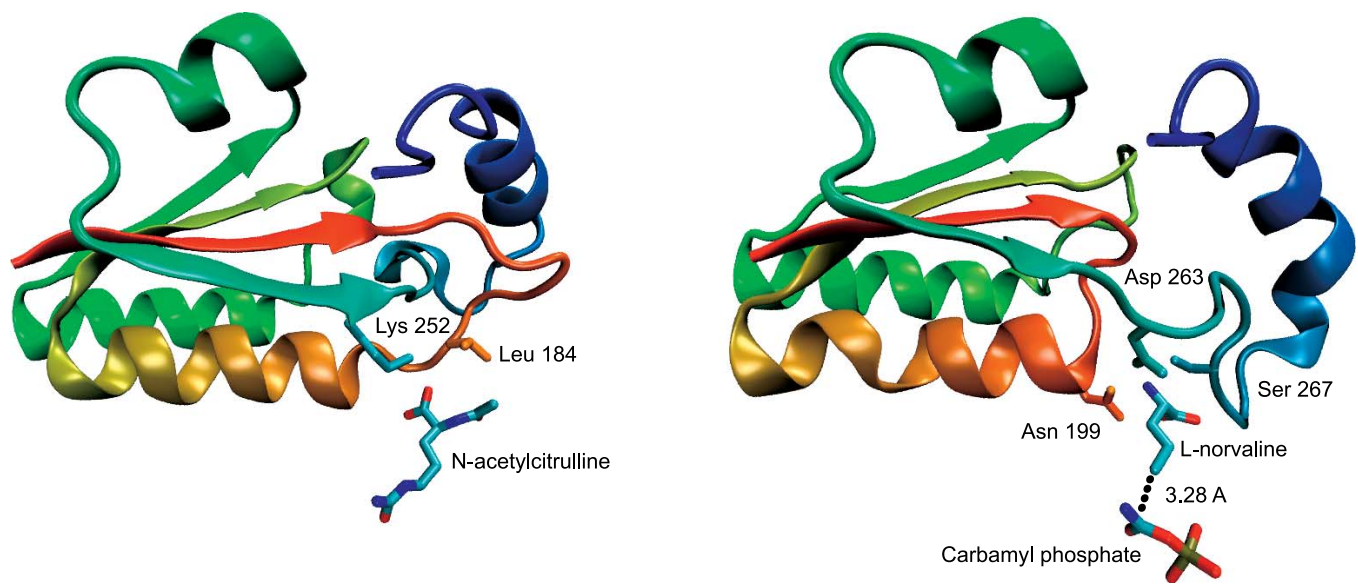

**Figure 2.** Structures of Transcarbamylase from *X. campestris* with a Trefoil Knot and from *Human* without a Knot
(Left) Knotted section (residues 171–278) of N-acetylornithine transcarbamylase from *X. campestris* with reaction product N-acetylcitrulline (pdb code 1yh1 [17]) and interacting side chains.
(Right) Corresponding (unknotted) section (residues 189–286) in human ornithine transcarbamylase (pdb code 1c9y [31]) with inhibitor L-norvaline and carbamyl phosphate. Colors change continuously from red (first residue in the section) to blue (last residue in the section). The two proteins have an overall sequence identity of 29% [41]. Pictures were generated with VMD [43].
DOI: 10.1371/journal.pcbi.0020122.g002

UCH-L3 in human and yeast share only 33% [29] of their sequences, but contain the same 5-fold knot as far as we can tell from the incomplete structure in yeast. It is not only likely that all species in between have the same knot—the link between sequence and structure may also be used to predict candidates for knots among isozymes or related proteins for which the structure is unknown. For example, UCH-L4 in mouse has 96% sequence identity with human UCH-L3. The similarity with UCH-L6 in chicken is 86%, and with UCH-L1 about 55%. Indeed, a reexamination of the most recent Protein Data Bank entries revealed that UCH-L1 contains the same $5_2$ knot as UCH-L3. (See the Update section—the structure was not yet part of the January Protein Data Bank release on which this paper is based.) Unfortunately, the method is not foolproof because differences between knotted and unknotted structures are sometime subtle. As we will demonstrate in the next paragraph, a more reliable estimate has to consider the conservation of major elements of the knot, like loops and threads.

**AOTCase—How a protein knot can alter enzymatic activity.** Somewhat surprisingly, we also identified a pair of homologues for which topology is not preserved. N-acetylornithine transcarbamylase (AOTCase [17]) is essential for the arginine biosynthesis in several major pathogens. In other bacteria, animals, and humans, a homologous enzyme (OTCase) processes L-ornithine instead [30]. Both proteins have two active sites. The first one binds carbamyl phosphate to the enzyme. The second site binds acetylornithine in AOTCases and L-ornithine in OTCases, enabling a reaction with carbamyl phosphate to form acetylcitrulline or citrulline, respectively [17, 31].

AOTCase in *X. campestris* has 41% sequence identity with OTCase from *Pyrococcus furiosus* [32] and 29% with human OTCase [31]. As demonstrated in Figure 2, AOTCase contains a deep trefoil knot which is not present in OTCase (Figure 2, right) and which modifies the second active site. The knot consists of a rigid proline-rich loop (residues 178–185), through which residues 252 to 256 are threaded and affixed. As elaborated in [17], the reaction product N-acetylcitrulline strongly interacts with the loop and with $Lys^{252}$. Access to subsequent residues is, however, restricted by the knot. L-norvaline in Figure 2 (right) is very similar to L-ornithine but lacks the N-ε atom of the latter to prevent a reaction with carbamyl phosphate. As the knot is not present in OTCase, the ligand has complete access to the dangling residues 263–268 and strongly interacts with them [31]. This leads to a rotation of the carboxyl-group by roughly 110° around the $C_\alpha$–$C_\beta$ bond [17].

This example demonstrates how the presence of a knot can modify active sites and alter the enzymatic activity of a protein—in this case, from processing L-ornithine to N-acetyl-L-ornithine. It is also easy to imagine how this alteration happened: a short insertion extends the loop and modifies the folding pathway of the protein.

## Discussion

Nature appears to disfavour entanglements, and evolution has developed mechanisms to avoid knots. Human DNA wraps around histone proteins, and the rigidity of DNA allows it to form a spool when it is fed into a viral capsid. One end also stays in the loading channel and prevents subsequent equilibration [33]. Knotted proteins are rare, although the reason is far less well understood. Can the absence of entanglement be explained in terms of particular statistical ensembles, or is there an evolutionary bias? And how do these structures actually fold?

Knots are ubiquitous in globular homopolymers [1–3,8], but rare in coil-like phases [1,34–36]. It is likely that even a flexible polymer will at least initially remain unknotted after a collapse from a swollen state. In proteins, the free energy landscape is considerably more complex, which may allow most proteins to stay unknotted. The secondary structure and the stiffness of the protein backbone may shift the length scale at which knots typically appear, too [8]. If knotted proteins are in fact more difficult to degrade, it might also be disadvantageous for most proteins to be knotted in the first place.

Unfortunately, few experimental papers address folding and biophysical aspects of knots in proteins. In recent work [37], Jackson and Mallam reversibly unfolded and folded a knotted methyltransferase in vitro, indicating that chaperones are not a necessary prerequisite. In a subsequent study [38], the authors provide an extensive kinetic analysis of the folding pathway. In conclusion, we would like to express our hope that this report will inspire more experiments in this small but nevertheless fascinating field.

## Materials and Methods

To determine whether a structure is knotted, we reduce the protein to its backbone, and draw two lines outward starting at the termini in the direction of the connection line between the center of mass of the backbone and the respective ends. These two lines are joined by a big loop, and the structure is classified by the determination of its Alexander polynomial [1,13]. To determine the size of the knotted core, we delete successively amino acids from the N-terminus until the protein becomes unknotted [1,6]. The procedure is repeated at the C-terminus starting with the last structure that contained the original knot. For each deletion, the outward pointing line through the new termini is parallel to the respective lines computed for the full structure. The thus determined size should, however, only be regarded as a guideline. A better estimate can be achieved by looking at the structure.

In Table 1 we include knotted structures with no missing amino acids in the center of the protein. (A list of potentially knotted structures with missing amino acids can be found in Table S3.) Technically, the numbering of the residues in the mmcif file has to be subsequent, and no two amino acids are allowed to be more than 6 Å apart. In addition, the knot has to persist when two amino acids are cut from either terminus. We have further excluded structures for which unknotted counterexamples exist (e.g., only one nuclear magnetic resonance structure among many is knotted or another structure of the same protein is unknotted). If a structure is fragmented, the knot has to appear in one fragment and in the resulting structure obtained from connecting missing sections by straight lines. Other knotted structures are only considered when at least one additional member of the same structural family [9] contains a knot according to the criteria above.

The enforcement of these rules leads to the exclusion of the bluetongue virus core protein [6] ($4_1$) and photoreceptor phytochrome A in *D. radiodurans* [11] ($3_1$), which have been previously identified as being knotted. Both structures are fragmented and become knotted only when a few missing fragments are connected by straight lines. In the viral core protein, the dangling C-terminus threads through a loose loop and becomes knotted in one out of two cases. On the other hand, the photoreceptor phytochrome A appears to contain a true knot. Notably, our analysis suggests that the thus connected structure of phytochrome A contains a figure-eight knot instead of a trefoil as reported in [11]. Moreover, we excluded a structure of the *Autographa California nuclear polyhedrosis virus*, which contains a knot according to our criteria. However, the N-terminus is buried inside the protein and the knot only exists because of our specific connection to the outside.

To further validate our criteria, we implemented an alternative method [4,8,39] that relies on the statistical analysis of multiple random closures. We arbitrarily chose two points on a sphere (which has to be larger than the protein) and connected each with one terminus. The two points can be joined unambiguously, and the resulting loop was analyzed by calculating the Alexander polynomial. We repeated the procedure 1,000 times, and defined the knot as the majority type.

Applying this analysis, we discovered 241 knotted structures in the Protein Data Bank. All 241 structures are also present in the 273 structures (Table S1) that were identified by our method, and the knot type is the same. The missing 32 structures (Table S2) are mostly shallow knots and were already rejected according to our extended criteria. The random closure also correctly discards rare structures with buried termini. In conclusion, the method used in this paper is considerably faster but requires a slightly increased inspection effort. Our observations agree with [8], which provides an extensive comparison of closures applied to proteins. A complete listing of knotted Protein Data Bank structures is given in the Supporting Information.

**Update.** Recently, the structure of human UCH-L1 was solved and released [40]. The protein shares 55% sequence identity with UCH-L3 [41], and it contains the same 5-fold knot. UCH-L1 is highly abundant in the brain, comprising up to 2% of the total brain protein [42]. The structure of UCH-L1 was not yet part of the January Protein Data Bank edition on which the rest of this study is based. We also noticed several new structures of knotted transcarbamylase-like proteins.

## Supporting Information

**Table S1.** List of Knotted Protein Data Bank Entries

Found at DOI: 10.1371/journal.pcbi.0020122.st001 (79 KB DOC).

**Table S2.** List of Knotted Entries from Table S1 That Become Unknotted When Ends Are Connected by the Random Closure Method

Found at DOI: 10.1371/journal.pcbi.0020122.st002 (28 KB DOC).

**Table S3.** List of Structures That Become Knotted When Missing Sections Are Joined by Straight Lines

Found at DOI: 10.1371/journal.pcbi.0020122.st003 (35 KB DOC).

### Accession Numbers

The Protein Data Bank (http://www.pdb.org) accession numbers for the structures discussed in this paper are human *UCH-L3* (1xd3), *UCH-L3* yeast homologue (1cmx), human *UCH-L1* (2etl), photoreceptor phytochrome A in *D. radiodurans* (1ztu), class II ketol-acid reductoisomerase in *E. coli* (1yrl), class II ketol-acid reductoisomerase in spinach (1yve), S-adenosylmethione synthetase in *E. coli* (1fug), S-adenosylmethione synthetase in rat (1qm4), AOTCase from *X. campestris* (1yh1), SOTCase from *B. fragilis* (1js1), OTCase from *P. furiosus* (1a1s), OTCase from human (1c9y), bluetongue virus core protein (2btv), and baculovirus P35 protein in *Autographa California nuclear polyhedrosis virus* (1p35).

## Acknowledgments

Upon completion of this work we became aware of a related study [8], which independently identified the knots in *UCH-L3* and *SOTCase* in a re-examination of protein knots. PV would like to acknowledge discussions with François Nédélec and with Olav Zimmermann, in which they proposed the potential link between protein knots and degradation. LM and PV would also like to thank Rachel Gaudet for a discussion about the function of ubiquitin hydrolase.

**Author contributions.** MK conceived the study. PV designed and wrote the analysis code. PV and LM analyzed the data. PV, LM, and MK wrote the paper.

**Funding.** This work was supported by National Science Foundation grant DMR-04–26677 and by Deutsche Forschungsgemeinschaft grant VI 237/1. LM is an Alfred P. Sloan Research Fellow.

**Competing interests.** The authors have declared that no competing interests exist.